\documentstyle[prd,aps]{revtex}
\begin{document}
\draft
\twocolumn[\hsize\textwidth\columnwidth\hsize\csname
@twocolumnfalse\endcsname
\preprint{SU-ITP-96-35    ~~~~~July 13, 1996 ~~~~ hep-th/9607108}
\title{  Moduli, Scalar Charges, and the   First Law of Black Hole
Thermodynamics}
\author{\bf Gary  Gibbons}\address{DAMTP,  Cambridge University, Silver Street,
Cambridge
CB3 9EW,
United Kingdom} \address{E-mail: G.W.Gibbons@damtp.cam.ac.uk}
\author{\bf Renata
Kallosh}
\address{ Department of Physics, Stanford University, Stanford, CA
94305, USA} \address{E-mail: kallosh@renata.stanford.edu}
\author{\bf Barak  Kol}
\address{ Department of Physics, Stanford University, Stanford, CA
94305, USA} \address{E-mail: barak@leland.stanford.edu}
\date{July 23, 1996}
\maketitle
\begin{abstract}
We show that under variation of moduli fields $\phi$ the first law of
black
hole thermodynamics becomes $dM = {\kappa dA\over 8\pi} + \Omega dJ +
\psi dq +
\chi dp - \Sigma d\phi$,
where $\Sigma$ are the scalar charges. We also show that the ADM mass
is
extremized at fixed
$A$, $J$,  $(p,q)$ when the moduli fields take the fixed  value
$\phi_{\rm
fix}(p,q)$ which depend only on electric and magnetic charges. It
follows that
the least mass of any black hole with fixed conserved electric and
magnetic
charges is given by the mass of the double-extreme black hole with
these
charges. Our work allows us to interpret the previously established
result
that for all extreme black holes the moduli fields at the horizon
take a value
$\phi= \phi_{\rm fix}(p,q)$ depending only on the electric and
magnetic
conserved charges:  $ \phi_{\rm fix}(p,q)$  is such that the scalar
charges
$\Sigma ( \phi_{\rm fix}, (p,q))=0$.
\end{abstract}
\pacs{PACS\,  04.70.Dy, 11.30.Pb, 11.25.-w  \hskip 2 cm SU-ITP-96-35
\hskip 2.5
cm hep-th/9607108}
\vskip2pc]

There has recently been intense interest in the thermodynamics of
black holes
in string theory. In particular the entropy $S$ of some extreme black
holes
considered as a function of their conserved electric and magnetic
charges
$(p,q)$ has been related to the logarithm of the number of the BPS
states at
large $(p,q)$ \cite{SV}. The properties of the black holes in the
theories
considered depend on the values $\phi_{\infty}$ of certain massless
scalar
fields, referred to as moduli fields, at spatial infinity. The moduli
at
infinity $\phi_{\infty}$ may be thought of as labelling different
ground states
or vacua of the theory. It is of crucial importance for the
consistency of the
state counting interpretation that the entropy $S={1\over 4}A$, where
$A$ is
the area  of the event horizon, is independent (in the extreme limit)
of the
particular vacuum or ground state, i.e. of $\phi_{\infty}$\,, and
depends only
on
the conserved charges $(p,q)$. The ADM mass $M$, however, does depend
on
$\phi_{\infty}$ even in the extreme case. In the non-extreme case
both the mass
$M$ and the area $A$ depend in a non-trivial way on $\phi_{\infty}$.
In other
words, to specify completely a black hole in these theories one needs
to give
the entropy
$S={1\over 4}A$, the conserved charges $(p,q)$, moduli at infinity
$\phi_{\infty}$, and the total angular momentum $J$.
In thermodynamic terms $A, (p^{\Lambda}, q_{\Lambda}), J,
\phi^a_{\infty}$ are
coordinates on the state space
${\bf R}_+ \times {\bf R}^{2n} \times {\bf R} \times {\cal
M}_{\phi}$, where
$\Lambda= 1, \dots , n$ is the number of electric (or magnetic)
charges, and
${\cal M}_{\phi}$ is the manifold in which the scalars take their
values, and
$a= 1, \dots , m= {\rm dim} {\cal M}_{\phi}$.

The usual first law of thermodynamics relates the variation of $M$ to
the
temperature $T= {\kappa \over 2\pi}$, where $\kappa$ is the surface
gravity,
the angular velocity $\Omega$ and the electrostatic  and
magnetostatic
potentials $\psi^\Lambda $ and $\chi_\Lambda$:
\begin{equation}
dM = {\kappa dA\over 8\pi} + \Omega dJ + \psi^\Lambda dq_\Lambda +
\chi_\Lambda
dp^\Lambda  \ .
\label{1st}\end{equation}
However equation (\ref{1st}) does not take into account the
dependence upon the
moduli
$\phi_{\infty}$. It should clearly be replaced by
\begin{equation}
dM = {\kappa dA\over 8\pi} + \Omega dJ + \psi^\Lambda dq_\Lambda +
\chi_\Lambda
dp^\Lambda + \left( {\partial M \over \partial \phi^a}\right)
d\phi^a \ ,
\label{1st+}
\end{equation}
where the partial derivative of the mass is taken at fixed values of
the area,
angular momentum and charges: $\left( {\partial M \over \partial
\phi^a}\right)_{A, J, p,q} $.

Our first result is that the coefficient of $d\phi^a$ is given by
\begin{equation}
\left( {\partial M \over \partial \phi^a}\right)_{A, J, p,q} = -
G_{ab}(\phi_{\infty}) \Sigma ^b\ ,
\label{scalarcharge}\end{equation}
where $G_{ab}$ is the metric on the scalar manifold ${\cal M}_{\phi}$
in terms
of which the kinetic part of the scalar Lagrangian density is
\begin{equation}
\frac{1}{2}
G_{ab} \partial_ \mu \phi ^a  \partial_\nu \phi ^b g^{\mu\nu}
\sqrt{-g} \ ,
\label{scalaraction}\end{equation}
and $\Sigma ^a$ are the scalar charges of the black hole defined by
\begin{equation}
\phi^a = \phi^a_{\infty} + {\Sigma^a \over r} + O({1\over r^2})
\end{equation}
at spatial infinity. Note that the scalar charges $\Sigma ^a$
themselves depend
non-trivially
on $A, (p^{\Lambda}, q_{\Lambda}), J, \phi^a_{\infty}$.
The vector part of the Lagrangian is
\begin{equation}
-\frac{1}{4}
 (\mu_{\Lambda\Sigma} {\cal F}^{\Lambda} {\cal
F}^{\Sigma}
- \nu_{\Lambda\Sigma} {\cal F}^{\Lambda}{}^*{\cal
F}^{\Sigma})
 \sqrt{-g} \ ,
\label{vectoraction}\end{equation}
where  the abelian field strengths are ${\cal F}^{\Lambda} \equiv
\partial
_{\mu} A_{\nu}^{\Lambda} -\partial _{\nu} A_{\mu}^{\Lambda}$ and
${}^*{\cal
F}^{\Sigma}
$ are the dual field strengths of the vector fields and
$\mu_{\Lambda\Sigma}$
and  $\nu_{\Lambda\Sigma}$ are
moduli dependent $n\times n$ matrices.  The charges
$(q_\Lambda,p^\Lambda)$
are defined by
\begin{eqnarray}\label{qqq}
p^\Lambda &=& {1\over 4\pi}\int {\cal
F}^\Lambda\ , \nonumber \\
q_\Lambda &=& {1\over 4\pi}\int (\mu_{\Lambda \Sigma}* {\cal
F}^ \Sigma + \nu_{\Lambda \Sigma} {\cal
F}^ \Sigma)\ .
\end{eqnarray}
We would like to stress that the charges must be defined as above, in
order
that  Gauss's theorem holds, i.e. the charges are
conserved and are the subject to quantization conditions in the
quantum theory.

One may prove eqs. (\ref{1st+}), with  (\ref{scalarcharge}) and
(\ref{scalaraction}),  either using Hamiltonian methods, modifying
the
procedure of Wald  \cite{Wald}, or by covariant methods, following
the older
procedure of Bardeen, Carter and Hawking \cite{BCH}. A recent account
of the
covariant approach including scalars but dropping the last terms of
eq.
(\ref{1st+}) is given in \cite{HS}.  From eq. (95) of \cite{HS}  for
gravity
coupled to a $\sigma$-model we have:
\begin{equation}
dM -  {\kappa dA\over 8\pi} - \Omega dJ =- \oint{d\phi^a G_{ab}(\phi)
{\partial \phi^b \over \partial x^i } d\sigma^i }\ ,
\label{deriv}\end{equation}
where the integral on the right hand side is over the boundary of a
spacelike
surface.
The boundary has two components, one on the horizon and one at
spatial
infinity. The contribution from the horizon vanishes because $\phi^b$
is
assumed to be independent of time and regular. The term at infinity
yields
\begin{equation}
dM -  {\kappa dA\over 8\pi} - \Omega dJ =- \Sigma^a  G_{ab}  d\phi^b
\ .
\label{derivation}\end{equation}
If vectors are present there is the usual additional term due to
variation of
the charges.

The last term in eq. (\ref{derivation}) was dropped in \cite{HS}
because in the application the authors had in mind  (Skyrmion black
hole) the
scalar charges $\Sigma ^a$ do indeed vanish.

For black holes in string theory, however,  the scalar charges
$\Sigma ^a$
will
not in general vanish. They will vanish if and only if
$\phi_{\infty}$, and
hence the vacuum state, is chosen to extremize the ADM mass at the
fixed
entropy ${A\over 4}$, angular momentum $J$, and conserved electric
and magnetic
charges $(p^{\Lambda}, q_{\Lambda})$. Note that despite the extra
term in the
first law the integrated version, i.e. the Smarr formula, remains
\cite{BGM}
\begin{equation}
M = {\kappa A\over 4\pi} +2\Omega J + \psi^\Lambda q_\Lambda +
\chi_\Lambda
p^\Lambda  \ .
\label{1st000}\end{equation}
{}From now on we will,
for simplicity, consider only static non-rotating black holes. The
extension to
include rotation is
both obvious and immediate.

The idea of extremization of the black hole mass in the moduli space
at the
fixed charges
was suggested for supersymmetric black holes by Ferrara and one of
the authors
\cite{FK}.
This idea is extended here for general black holes.

Our second result is that subject to a convexity condition that we
explain
below, the scalar charges vanish and hence $M$ is extremal if and
only if the
black hole solution has constant values of the moduli fields
\begin{equation}
\phi^a (x)= \phi^a_{\infty} \ .
\end{equation}
Moreover the constant values  $\phi^a_{\infty}$ is not arbitrary  but
must be
chosen to extremize
at fixed electric and magnetic charges a certain non-negative
function $V$
which is quadratic in the electric and magnetic charges and depends
non-trivially on the scalars.
\begin{equation}
V= (p,q)^t   {\cal M}  \left (\matrix{
p\cr
q\cr
}\right ),
\end{equation}
where
\begin{equation}
 {\cal M}^{-1} = \left |\matrix{
\mu+ \nu \mu^{-1} \nu  & \nu \mu^{-1}  \cr
\mu^{-1} \nu &  \mu^{-1} \cr
}\right |.
\end{equation}
 For extended supergravity  theories these
$2n\times 2n$ moduli dependent matrices have been studied before
\cite{BGM}, \cite{FK}, \cite{KSW}.

Spherically symmetric non-extreme black holes in theories described
above can
be conveniently cast into the form \cite{Gib82}
\begin{eqnarray}
ds^2 &&= - e^{2U} dt^2 + e^{-2U} \Bigl[ {c^2 d\tau^2 \over \sinh^4 c\tau}
+ {c^2
\over \sinh^2 c\tau} \nonumber\\
&&(d \theta ^2 +  \sin^2 \theta d \varphi ^2)\Bigr] \ .
\end{eqnarray}
The coordinate $\tau$ runs from $-\infty$ (horizon) to $0$ (spatial
infinity).
The boundary condition for $U$ is that  $U(0)=1$ and $U \rightarrow
c\tau$
as $\tau \rightarrow -\infty$. The boundary condition for
$\phi^a(\tau)$ is
that $\phi^a(0)=\phi^a_{\infty} $ and ${d\phi^a \over d\tau }=
O(e^{c\tau})$ as
$\tau \rightarrow -\infty$. The physical significance of $c$ is that
\begin{equation}
c= {\kappa A\over 4\pi} = 2ST \ .
\end{equation}
The field equations for $U$ and $\phi^a$ are
\begin{eqnarray}
{d^2 U \over d\tau^2} &=& 2 V(\phi, (p,q)) e^{2U}, \\
{D\phi^a \over D \tau^2} &=& {\partial V \over \partial \phi^a}
e^{2U},
\end{eqnarray}
and
\begin{equation}
\left ({d U \over d\tau}\right )^2 + G_{ab}  {d\phi^a \over d\tau }
{d\phi^b
\over d\tau } -
V(\phi, (p,q)) e^{2U} = c^2\ .
\label{Ueq} \end{equation}

 Our convexity condition is that  the symmetric tensor field on
${\cal
M}_{\phi}$ defined by
\begin{equation}
V_{ab} = \nabla_a  \nabla_b V \ ,
\end{equation}
where $ \nabla_a$ is the Levi-Civita covariant derivative with
respect to the
metric $G_{ab}$ of
${\cal M}_{\phi}$\,, is non-negative.

It follows from the equation of motion for $\phi ^a$ that
\begin{equation}
{d^2 V \over {d\tau ^2} } = V_{ab}  {d\phi^a \over d\tau } {d\phi^b
\over d\tau
} +
2 e^{2U} {\partial V \over \partial \phi^a} {\partial V \over
\partial \phi^b}
G^{ab} \ .
\end{equation}

If we multiply by V, integrate and use the boundary conditions, we
obtain
\begin{eqnarray}
&-& \int_{-\infty}^{0} {{1 \over 2} \left( {dV \over d\tau} \right)^2
d\tau} =
\Sigma ^a \left(  {\partial V \over \partial \phi^a} \right)_\infty
+\nonumber\\
&+& \int_{-\infty}^{0}{\left( V_{ab}  {d\phi^a \over d\tau } {d\phi^b
\over
d\tau } +
2 e^{2U} {\partial V \over \partial \phi^a} {\partial V \over
\partial \phi^b}
G^{ab} \right)} d\tau \ .
\end{eqnarray}

If we assume that $\Sigma ^a =0$ and $V_{ab}$ is positive definite,
we must
have ${\partial V \over \partial \phi^a } =0$ for all $\tau$, which
implies that
the moduli are frozen, i.e. $\phi^a (r)=\phi^a_\infty$.

The mass of the black hole is given by $M= ({dU\over d\tau}) _{\tau=0}$ and
therefore
we have
from (\ref{Ueq}) a  rather useful general relation \cite{Gib82} which may be
interpreted as the statement that the total self force on the
  hole due to the attractive forces of gravity and the scalar fields
is not exceeded by the repulsive self force due to the vectors
and vanishes only in the extreme case:
\begin{equation}
M^2 + G_{ab} \Sigma ^a \Sigma ^b - V(\phi^a_\infty) = 4 S^2 T^2 \ .
\end{equation}
One might refer to the inequality obtained from the non-negativity of the right
hand side  as an antigravity bound. Note that unlike the Bogomol'nyi bound
\cite{GH} its derivation requires neither
supersymmetry nor duality invariance.
Differentiating with respect to $\phi^a_\infty$ gives:
\begin{equation}
M {\partial M \over \partial \phi^c_\infty} + G_{ab} \Sigma ^a \nabla
_c \Sigma
^b - {1 \over 2} {\partial V \over \partial \phi^c_\infty} = 4 S^2 T
{\partial T
\over \partial \phi^c_\infty}\ .
\end{equation}

We deduce that if the mass is extremized with respect to
$\phi^a_\infty$, then
so is the temperature.
It follows that the fixed or ``frozen"
moduli must minimize $V$, i.e. $\phi_{\rm fix}$ is defined by
\begin{equation}
\left({\partial V \over \partial \phi^a}\right)_{\phi= \phi_{\rm
fix}, (p,q)}
=0 \ .
\end{equation}

Static black holes with frozen moduli have the space-time geometry
given by the
Reissner-Nordstr\"{o}m metric.

A year ago Ferrara, Kallosh, and Strominger \cite{FKS} found that for
a class
of
supersymmetric black holes the moduli field at the horizon,
$\phi_{H}$, depends
only on the conserved electric and magnetic charges
\begin{equation}
\phi_{H, \rm {extreme}} = \phi_{\rm fix} (p,q) \ .
\end{equation}
Recall that at extremality, the mass depends on the moduli at
infinity and the
conserved charges
\begin{equation}
M= M_{\rm {extreme}} (\phi^a_{\infty}, (p,q)) \ .
\end{equation}
An implicit formula was found more recently in \cite{FK} for
$\phi_{H, \rm
{extreme}}$  that can be written
as
\begin{equation}
\left({\partial M_{\rm {extreme}} \over \partial \phi}\right)_{
(p,q), \phi=
\phi_{H, \rm {extreme}}} =0 \ ,
\label{Fer}\end{equation}
where the derivative is taken at fixed values of charges.
The result which holds for all of the theories we consider here was
found by
analyzing the radial equation for the moduli fields $\phi(r)$ which
is governed
by the function $V(\phi, p,q)$.

Two questions arose and motivated the results of this paper:
\begin{itemize}
\item  Why is $\phi_{H, \rm {extreme}}$ independent of $\phi_{\infty}
$?
\item Why is $\phi_{H, \rm {extreme}}$ given by (\ref{Fer})?
\end{itemize}

We can now offer an answer for the second question. From
(\ref{scalarcharge}),
which we first derived for the example given in \cite{KLOPP}, it
follows that
equation (\ref{Fer}) is equivalent to
\begin{equation}
\Sigma^a ( \phi_{\rm fix},  (p,q))
=0 \ ,
\label{barak}\end{equation}
thus defining $ \phi_{\rm fix}= \phi_{\rm fix}(p,q)$. (This equation
was noted
in \cite{FK} for the extreme case).
But as we stated above a black hole with vanishing scalar charge must
have
spatially constant moduli fields : $\phi^a (r) = \phi^a _{H, \rm
{extreme}} =
\phi^a_{\infty}$, or ``frozen" moduli. In other words, to satisfy eq.
(\ref{barak})  we must choose $\phi^a_{\infty}$ to be $\phi^a _{H,
\rm
{extreme}}$.

As found in \cite{FK}, the entropy of  all extreme black holes is
independent
of $\phi^a_{\infty}$
and is given by
\begin{equation}
S = {A\over 4} = \pi V\left( \phi_{\rm fix} (p,q), (p,q)\right) \ .
\end{equation}

Our new result establishes that for any static black hole, extreme or
not
\begin{equation}
M(S, \phi_{\infty}, (p,q)) \geq M(S, \phi_{\rm fix}, (p,q)) \ .
\end{equation}
But because black holes with frozen moduli
have the Reissner-Nordstr\"{o}m geometry, the right hand side of (16)
is always
greater than the mass of the extreme Reissner-Nordstr\"{o}m black
hole with
same
charges.

We  would like to
emphasize that our results hold for a wide class of theories -- one
need not
assume either supersymmetry or duality invariance. In addition we
wish to
emphasize the following:

i) The   scalar charges $\Sigma^a$ are not conserved but they do act
as the
sources for the moduli.  They are not associated with a conserved
current. The
flux of the gradient of the scalar charge vanishes at the horizon.
Thus the
scalar charge resides entirely outside the event horizon.

ii) Previously one did not consider variations of the moduli at
infinity,
$\phi_{\infty}$\,, which were regarded fixed once and for all. In
that case the
scalar charge $\Sigma^a$ need not be specified independently of the
mass,
angular momentum and electric and magnetic charges. However if one
does not
regard the moduli at infinity to be given {\it a priori} one needs to
specify,
in addition to M, J and (q,p), either $\phi_{\infty} $ or  $\Sigma^a$
to
characterize completely the black hole. This may be important when
considering
situations in which  $\phi_{\infty} $ becomes dynamical, for example
if one
considers slow adiabatic changes of   $\phi_{\infty}$  or possibly
time-dependent cosmological situations.

\

 Stimulating discussions with S. Ferrara, A. Linde,  A. Peet, A.
Rajaraman,  L. Susskind,
and W. K. Wong are gratefully acknowledged. Barak thanks his teacher
L. Susskind, Arvind
Rajaraman and Edi Halyo. GWG would like to thank the  Stanford
Theoretical
Physics
Institute for the hospitality and support during the period of this
work.
This work is supported by the  NSF grant PHY-9219345.

\end{document}